# A multi beam proton accelerator

S. N. Dolya

*Joint Institute for Nuclear Research, Joliot - Curie 6, Dubna, Russia, 141980*

**Abstract**

The article considers a proton accelerator containing 7 independent beams arranged on the accelerator radius. The current in each beam is Ip = 0.1 A. The initial part of the accelerator consists of shielded spiral waveguides assembled in the common screen. The frequency of the acceleration f = 300 MHz, high-frequency power P = 25 MW, the length of the accelerator $L_{acc1}$ = 6 m. After reaching the proton energy of $W_{p1}$ = 6 MeV the protons using lenses with the azimuthal magnetic field are collected in one beam. Further beam acceleration is performed in the array of superconducting cavities tuned to the frequency f = 1.3 GHz. The acceleration rate is equal to 20 MeV / m, the high-frequency power consumption P = 15 MW / m.

**1. Introduction**

The problem of constructing a proton accelerator with the beam energy of about 1 GeV and the average beam current of $I_{av}$ ~ 1 mA consists of two sub-tasks. One of them is as follows: it requires a high rate of acceleration for to have the minimum length of the accelerator. Now the available rate of acceleration of protons equal to $\Delta W_p / \Delta l$ ~ 1 MeV / m is clearly not enough to solve the problem. We must strive to the acceleration rate of ~ 30 MeV / m, obtained in electron linear accelerators.

The difference of the proton accelerator from the electron one is as follows: the proton accelerator motion phase is not frozen, it is necessary to choose the synchronous phase in the range of $20^0$-$70^0$ and this synchronous phase is needed to change not very much during the passage of the bunch of the beam through the cavity. This will require from the superconducting cavities to be narrow enough in the direction of the proton acceleration. Let's start the acceleration of the proton energy of ~ 5 MeV, the proton velocity $v_p = 3 * 10^9$ cm / s, $\beta_p = 0.1$, where $\beta = v / c$, $c = 3*10^{10}$ cm / s which is the velocity of light in vacuum.

The length of the cavity along the axis of the acceleration can be estimated as $\beta\lambda/4$, where $\beta = 0.1$ is the initial proton velocity in this part of the accelerator, $\lambda$ - the wavelength of the acceleration, which is chosen: $\lambda = 2$ m. In our case $\beta\lambda/4 = 5$ cm, which can be considered to be acceptable. Before active using of superconductivity in accelerator technologies, the proton acceleration in the array of warm cavities was of no interest. It is explained by the following: to reach significant field tension in this array, it was required to have high-frequency power. This accelerating system turned out to be worse than other systems.

When superconducting cavities appeared the losses of high-frequency power in the cavities significantly reduced. Thus, it was possible to come back to the opportunity of using of this system for the acceleration of protons in it.

Another problem is related with the required high intensity of the accelerated proton beam. If we assume that the required average current of the accelerated beam is of the order of $I_{av} \approx 1$ mA, and the temporal structure of the accelerator is as follows: pulse duration $\tau_p = 200$ μs, the pulse repetition frequency $F = 5$ Hz, this means that the required pulse beam current is big: $I_{pp} \approx 1$ A.

It is easy to estimate the number of protons in a bunch for this case. The pulse intensity of protons corresponds to the number of particles $6 * 10^{18}$ p / s, and for the wave period $\tau_w = 6.7$ ns, the number of particles contained in a single bunch, is found to be equal to: $6 * 10^{18}$ p / s * 6.7 ns = $4 * 10^{10}$ protons per bunch.

The problem of preserving this amount of particles in one bunch will appear immediately after extraction of the non-modulated beam from the ion source. For the modulated beam the situation looks just catastrophic. Assume the beam radius to be equal to $r_b = 0.4$ cm.

If we assume that the source is under potential $U_{sou} = 50$ kV relatively the Earth, the initial velocity of the proton beam (from the source) is equal to $\beta_{in} = 10^{-2}$. This means that the length of the non-modulated beam part is equal to the wavelength $l_{bin} = \lambda_s = \lambda * \beta_{in} = 2$ cm. The modulated beam covers the phase interval approximately equal to $\Delta\varphi = 36^0$, i.e., the length of the bunch after the buncher will be equal to $l_{binb} = 2$ mm.

Thus the volume of the bunch is of the order of $\pi\, r_b^2 * l_{binb} \approx 0.1$ cm$^3$, and the bulk density of the particles in the bunch will be of the order of $n_p = 4 * 10^{11}$ p / cm$^3$.

To hold this bunch against spreading out along the radius of the bunch can be fulfilled with the solenoid magnetic field. Best of all the holding of the bunch is described by using the term "frequency".

If to consider that the ion source is beyond the magnetic field of the solenoid,

then getting into the magnetic field, the protons will spin in it with the Larmor frequency $\omega_L = eH / 2mc$, where $e = 5 * 10^{-10}$ –the elementary charge, $H \approx 10^5$ Gs - magnetic field coil, $m = 1.7 * 10^{-24}$ g - the rest mass of the proton. This frequency must be much larger than the plasma frequency:
$\omega_p = (4\pi e^2 n_p / m)^{1/2} = 1.3 * 10^3 * n_p^{1/2} = 8 * 10^8$ for selected parameters.

Substituting numbers in the expression for the Larmor frequency, we find that the frequency is equal to $\omega_L = 5 * 10^8$, and it is less than the plasma frequency. This means that this magnetic field in the chosen beam size will not be able to hold the bunch over the radial direction. It is necessary to increase the voltage at which the source is located relatively the Earth. If the voltage is increased by 16 times to U = 800 kV, then the beam velocity will increase by 4 times. The bunch length will increase by 4 times, and the proton density in the bunch will also drop by 4 times. The plasma frequency will reduce by twice till the value of $\omega_{p1} \approx 4 * 10^8$. This density is possible to keep in the radial direction by the magnetic superconducting solenoid field.

Now let us consider the issue whether it is possible to keep this proton density in the bunch in the longitudinal direction. We will follow [1] while considering this issue.

We introduce parameter $W_\lambda$ which is a specific acceleration equal to the ratio of the proton energy addition for this wavelength to the rest energy of the proton. For the traveling wave this parameter is equal to the following:

$$W_\lambda = eE\lambda * \cos\varphi_s / mc^2, \qquad (1)$$

where E – the tension of the accelerating electric field, $\varphi_s$ is the synchronous phase. We choose the synchronous phase equal to: $\varphi_s = 30^0$. This is usually a small parameter in proton accelerators. Indeed, substituting E = 20 kV / cm, $\lambda = 2$ m, $\cos\varphi_s \approx 1$, $mc^2 = 1$ GeV, we obtain: $W_\lambda = 4 * 10^{-3}$.

The phase oscillation frequency is the frequency at which the protons oscillate in the bunch in the longitudinal direction:

$$\Omega_{ph} = \omega * (W_\lambda * tg\, \varphi_s / 2\pi\beta_s)^{1/2}, \qquad (2)$$

where $\omega = 2\pi f$ is the circular frequency of the proton acceleration. Substituting the numbers into the formula (2) we obtain, $\Omega_{ph} \approx 0.1\, \omega$. It means that the frequency of the phase oscillations is less than the acceleration rate and it is $\Omega_{ph} \approx 10^8$. This frequency turned out to be lower than the plasma frequency: $\omega_{p1} = 4 * 10^8$. This demonstrates that to hold the bunch in the longitudinal direction by the wave field will not be possible in the required sizes. The bunch will spread out in the longitudinal direction.

The solution of this problem is in accelerating simultaneously several proton beams (seven) in one aperture. Then, by reducing the beam current in one beam till the value of $I_{p1} = 0.1$ A, it will be possible to receive the proton beam current pulse equal to: $I_p \approx 0.7$ A.

**2. Simultaneous formation of several proton beams in one ion source**

We assume that the ion source is seven duoplasmatrons assembled in one aperture. The distance between the centers of the beams is assumed to be equal to 6 cm. The current of each beam should be of the value of ~ 0.2 A, so that after bunching the proton beam with a factor of capture $k_k = 1/2$, the current value of the bunched beam was 0.1 A. By reducing the pulse current by the order of the magnitude the plasma frequency will reduce by 3 times, and it will become approximately equal to the frequency of the phase oscillations in the wave.

As for holding the proton beam in the radial direction, at the proton beam current $I_{pp} = 0.1$ A and the transverse beam dimension $r_b = 0.4$ cm it is possible to perform this holding by the longitudinal magnetic field with the intensity $H = 10$ T. The Larmor frequency of protons rotating in the magnetic field will be higher than the plasma oscillation frequency of protons in the bunch. It is important to emphasize that the solenoid magnetic field has no selected center and each of the seven beams will be held in the transverse direction relatively the initial velocity of the beam.

**3. Preliminary acceleration of the proton beams in spiral waveguides**

A spiral waveguide is theoretically a well-studied [2, 3] accelerating structure having a high rate of acceleration $\Delta W / \Delta l \sim 3$ MeV / m. The distinguished feature of the spiral waveguide is extremely small dimensions in the transverse direction.

We choose the initial radius of the spirals, where the proton beams will be accelerated, to be equal to: $r_{0in} = 1$ cm, the radius of the external screen will be chosen to be equal to:

$r_{sc} = 3$ cm. Then the whole assembly, consisting of seven spiral waveguides will have the outer diameter $D_{ass} = 18$ cm.

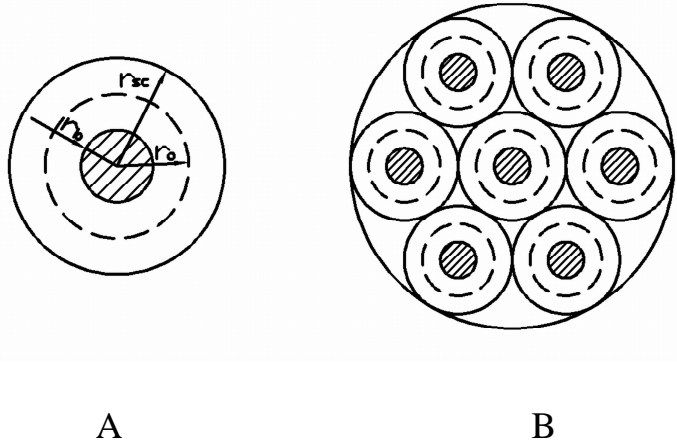

A                  B

Fig.1 The scheme of the beam location in the spiral waveguide (A) and assemble of the spiral waveguides (B).

It is possible to increase the frequency of the phase oscillations by means of selecting a synchronous phase sacrificing the acceleration rate. This way we increase the value of the accelerated current [4]. Indeed, if to select the synchronous phase to be equal to $\varphi_s = 60^0$, the rate of acceleration, formula (1), will drop by 1.7. The phase oscillation frequency will grow in the following relation: $\sin 60^0 / \sin 30^0$, by 1/2, i.e., - 1.3 times. It will result in increasing the phase oscillation frequency by 30%.

Not to carry out the calculations again, we will compare the conditions of acceleration in one of the seven modules with the parameters of the spiral waveguide discussed in [3].

Article [3] discussed the acceleration of ions of gold having the ratio of the charge $Z = 69$ to the atomic mass $A = 197$ equal to $Z / A = 0.3$. Protons having the ratio of the charge to the mass equal to: $Z / A = 1$ will reach the acceleration rate by 3 times greater at the selected acceleration parameters. This article [3] has selected the initial acceleration energy much smaller than $W_{in} = 0.8$ MeV, chosen for this given work. This energy in the accelerator of gold ions is achieved on the length of the acceleration equal to 1m. The parameters of the accelerator of the gold ions will be the same as for the proton accelerator after the length of acceleration equal to 1 m.

The initial radius of the spiral in [3] was chosen equal to 2 cm. We have chosen the initial radius of the spiral $r_{in}$ = 1 cm, i.e. by 2 times lower. We use the term for the spiral radius "initial" since it is assumed that the radius of the spiral winding will vary along the length of the accelerator reducing from the beginning spiral section to the end of the acceleration section.

This decreasing of the radius of spiral winding is conditioned by the following: while increasing the phase velocity of the wave the electric field tension reduces at winding on the cylindrical frame. The phase velocity of the wave should be continuously increased while ion accelerating. It is necessary to perform synchronization: the phase velocity of the waves to be kept equal to the velocity of the ions being accelerated.

The phase velocity of the wave in the spiral is approximately determined by the formula:

$$\beta_{ph} = \text{tg } \psi = h/2\pi r_0, \qquad (3)$$

where $\beta_{ph} = v_{ph}/c$ is the phase velocity, expressed in terms of the velocity of light, tg $\psi$ - tangent of spiral winding, h – the spiral winding step, $r_0$ - radius of the spiral winding. Formula (3) shows that if we reduced the winding radius of the spiral by 2 times, compared with the spiral discussed in [3], and at the same time we have increased by four times the phase velocity of the wave, then the radius of the spiral winding will remain exactly the same as considered in [3].

The spiral winding step can be increased significantly by using dielectric filling between the spiral and the shield, as suggested in [5]. It should, however, keep in mind that most of the flow of high-frequency power will be distributed over the dielectric, i.e., between the spiral and the screen. The power required to obtain the chosen electric field tension will increase significantly.

High-frequency power P is related with the electric field tension on the axis of the spiral $E_0$ by the following ratio [5]:

$$P = (c/8) E_0^2 r_0^2 [kk_3/k_1^2]\{(1+I_0K_1/I_1K_0)(I_1^2-I_0I_2) +$$

$$+ \varepsilon (I_0/K_0)^2(1+I_1K_0/I_0K_1)(K_0K_2-K_1^2)\}, \qquad (4)$$

where $k = \omega/c$ - wave vector, $k_1 = k(1/\beta_{ph}^2 - 1)^{1/2}$ - radial wave vector inside the spiral, $k_3 = \omega/v_{ph}$ - the wave vector in the axial direction, $I_0$, $I_1$, $I_2$ - modified functions by Bessel of the first type; $K_0$, $K_1$, $K_2$ are modified Bessel functions of the second type. The first term in the curly brackets corresponds to the flow of the power propagating inside the spiral; the second term corresponds to the flow of power propagating between the spiral and the screen. If this area is filled with dielectric with a dielectric constant $\varepsilon$, then before the second term there must be factor $\varepsilon$.

In the case of large slowing down, $\beta_{ph} \ll 1$, and the optimum ratio between the perimeter of the spiral turn, $2\pi r_0$, and the slowed down wavelength $\lambda_s = \beta_{ph} * \lambda$, where $\lambda = 2m$ that is the wavelength in vacuum, formula (4) is simplified. Under the optimum ratio we mean the ratio of $x = 2\pi r_0 / \beta_{ph} * \lambda \approx 1$. The expression $[kk_3 / k_1^2]$ can be replaced by $\beta_{ph}$. The value of the curled bracket for the argument of the Bessel functions where $x = 1$, is equal to $\{\} = 4.44$.

From the relation $x = 2\pi r_0 / \beta_{ph}*\lambda \approx 1$ it can be seen that the frequency $f = 150$ MHz is not optimal for the radius of the spiral $r_{0in} = 1$ cm. The reason of the above is that while accelerating the particles in a spiral waveguide the parameter x is permanently decreases. From one side, the spiral winding radius decreases: it is necessary to maintain the even acceleration rate along the section [5]. From the other side, the phase velocity of the wave increases during the acceleration of the protons in the section.

Table 1 shows the values of the first and second terms in the curly brackets in (4). We can see that with increasing of parameter x, the both terms, first, decrease monotonously. The sum of the both terms reaches the minimum of for the parameter value $x = 1$. It means the equality of the spiral turn perimeter and of the slowed down wavelength. The minimum value of the curled bracket in formula (4) means that at the given level of the high frequency power introduced into the spiral we will get the maximum field strength on the axis of the spiral.

Table 1. Values of the first and second terms in the curled bracket of formula (4).

| x   | I     | II    |
|-----|-------|-------|
| 0.1 | 0.1   | 66.8  |
| 0.2 | 0.14  | 22    |
| 0.3 | 0.18  | 12.14 |
| 0.4 | 0.226 | 8.286 |
| 0.5 | 0.273 | 6.365 |
| 0.6 | 0.326 | 5.277 |
| 0.7 | 0.386 | 4.618 |
| 0.8 | 0.454 | 4.208 |
| 0.9 | 0.532 | 3.958 |
| 1 | 0.620 | 3.819 |
| 1.1 | 0.721 | 3.763 |
| 1.2 | 0.836 | 3.774 |
| 1.3 | 0.968 | 3.844 |
| 1.4 | 1.119 | 3.96 |
| 1.5 | 1.29 | 4.142 |
| 1.6 | 1.494 | 4.369 |
| 1.7 | 1.724 | 4.650 |
| 1.8 | 1.989 | 4.989 |
| 1.9 | 2.295 | 5.393 |
| 2 | 2.69 | 5.867 |
| 2.5 | 5.441 | 9.68 |
| 3 | 11.336 | 17.601 |

At further increase of parameter x the numerical value of the curled bracket begins to increase monotonously. If to choose the value of x at the beginning of the acceleration slightly greater than 1, then the maximum tension of the electric accelerating field is reached in the middle of the section [5]. After that the field strength will start to diminish.

For the parameters: $x = 1.5$, the initial radius of the spiral $r_{0in} = 1$ cm, the initial velocity of the wave in the spiral, i.e., $\beta_{ph\ in} = 4 * 10^{-2}$, the optimal wavelength is $\lambda = 100$ cm, i.e., the frequency of acceleration $f = 300$ MHz. Note that the frequency of the phase oscillation increases proportional to the acceleration frequency according to the formula (2). Now the phase oscillation frequency exceeds the plasma frequency. It is obvious that the plasma frequency does not depend on the bunching frequency.

It is evident that the plasma frequency does not depend on bunching frequency. It is seen that if to compare with the work [3] - we have reduced the initial radius of the spiral by 2 times, and the initial phase velocity was increased 4 times. Then to achieve the same field tension $E = 20$ kV / cm, the same high-frequency power will be required: 3 MW. Since there are 7 modules in the assembly, then the total required high-frequency power is equal to 21 MW.

The high-frequency power of 21 MW will be used to form the required electric field tension. We must also take into account the proton beam influence on the proton acceleration of the section. We estimate it taking into consideration the following reasons. Let us assume that acceleration of the beam occurs in the continuous mode. This means that the power of 6 MV * 0.7A ≈ 4 MW is transmitted to the beam. The same power will be also transferred to the beam at the long ($\tau_p$ = 200 μs) acceleration pulse for the required high-frequency power should be equal to P ≈ 25 MW.

The length of the acceleration of protons up to 6 MeV can be estimated as follows. When the amplitude of the accelerating voltage E = 2 MV / m and the synchronous phase cosine cos $\varphi_s$ = ½, the rate of energy gain is equal to 1 MeV / m and the energy of 6 MeV will be obtained by the protons at the length of the accelerator equal to 6 m.

Here is a table of parameters of the initial part of the accelerator.

Table 2. Parameters of the initial part of the accelerator

| Option | Value |
|---|---|
| The voltage of the proton source is, kV | 800 |
| Pulse current of the proton source, mA | 7*0.2 |
| Bunched beam current, mA | 7*100 |
| The frequency of the acceleration f, MHz | 300 |
| The tension of the focusing magnetic field, T | 10 |
| The duration of the current pulse $\tau_b$, μs | 200 |
| The average tension of electric field $E_0$, MV /m | 2 |
| The cosine of the synchronous phase, cos$\varphi_s$ | 0.5 |
| The pulse repetition rate F, Hz | 5 |
| High-frequency power, MW | 25 |
| The length of the accelerator, m | 6 |

The protons will be influenced by the defocusing force from the waves. The frequency corresponding to this force is equal to: $\Omega_r = \Omega_{ph} / 1.41 = 7*10^7$. It is much smaller than the Larmor frequency: $\omega_L = 5 * 10^8$.

**4. Junction of the peripheral proton beams into a common beam.**

Independent acceleration of each of the seven beams at the initial stage of acceleration is required to provide the conditions to keep the accelerated bunch in the longitudinal and transverse directions. After reaching the velocity of about 0.1 by bunches it is possible to join all the seven beams into one.

Here is the sequence of operations required to form the proton beam of the thread form. The proton beams extracted from the spiral waveguides can be transferred to the radius close to the axis of proton acceleration after two turns of the proton beam by the same corner. At first, lens 1 with the azimuthal magnetic field turns the protons into the direction of the acceleration axis. When protons come to the acceleration axis their trajectory should be turned by lens 2 exactly by the same angle, but in the opposite direction.

Now we carry out the calculations of the proton beam trajectory.

Let protons gather onto the axis after 3 meters of passing through lens. This means that the radial angle which they acquire during the passage through the lens1 is equal to 6cm / 300cm = $2 * 10^{-2}$. Let $H_\varphi$ field in the lens is equal to $10^4$ Gs, the length of the track (along axis z) is $l_1 = 7$ cm, then for the protons having the rest energy equal to $mc^2 = 1$ GeV, the angle of the turn will be as follows:

$$\Theta = eH_\varphi * l_1/mc^2 = 10^4 * 3 * 10^2 * 7/10^9 = 2*10^{-2}, \qquad (5)$$

what is equal to the required angle. At a distance of 30 cm from the point where protons will cross the acceleration axis we embed the second lens exactly the same and using it we will turn the beam by the same angle but to the opposite direction. Thereafter, the beam will move parallel to the axis z, but at a radius in relation of 300/30 =10, i.e., by 10 times smaller. Thus, the radius of the proton beam will be equal to $r_b \approx 1.5$ cm. Exactly the same principle of compressing the broad beams is used in optics, where this is done by means of two lenses with different focal lengths.

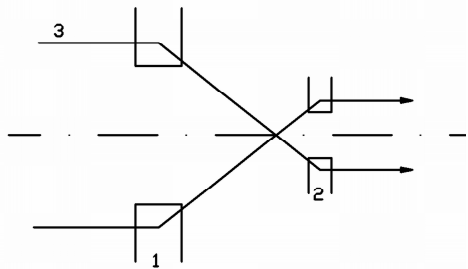

Fig. 2. Scheme of transferring the proton beams into the region near the acceleration axis. 1 –the first lens, 2 – the second lens, 3 – beam trajectories.

**5. Acceleration of protons in the joint of superconducting cavities**

After peripheral beams are joint to the acceleration axis in a common beam, it is necessary to demodulate it at the frequency f = 300 MHz, and after that – to modulate it at frequency $f_1$ = 1.3 GHz corresponding to the frequency of the international electronic linear collider [6].

The wavelength corresponding to the frequency $f_1$ = 1.3 GHz is approximately equal to $\lambda_1 \approx$ 24 cm. This means that at the excitation of the wave $E_{010}$ the diameter of the cavity will be about $d_{cav}$ = ¾ * $\lambda_1$ = 18 cm. The height of the cylindrical cavity, i.e., its length along the acceleration axis should be about $l_{in} = \beta_{in1} * \lambda_1 / 4 \approx$ 1 cm. The height of the cavity should increase with increasing the particle velocity from $\beta_{in1}$ = 0.1 at the beginning of acceleration to $\beta_{fin1}$ = 0.9 at the end of acceleration.

It is possible to calculate the quality factor of the cavity and the high frequency power required to achieve the given field strength in the warm cavity having the same dimensions as the superconducting one.

The formula that links the high-frequency power P with a voltage between the end walls of the cavity U, looks as follows:

$$P = U^2/2R_{sr}, \qquad (6)$$

where $R_{sr}$ - shunt resistance of the cavity. The shunt resistance of the cavity, in its turn, can be expressed in terms of the characteristic resistance $\rho_s$ and the quality factor Q. The shunt resistance of the cavity is equal to: $R_{sr} = \rho_s * Q$.

The quality factor Q of the cavity can be estimated as the ratio of the volume V of the cavity to the volume equal to the interior surface area multiplied by the depth of the skin - layer $\delta$. The depth of the skin- layer for copper at this frequency is equal to: $\delta = c / (2\pi\sigma\omega)^{1/2}$, where $\sigma = 5.4*10^{17}$ $s^{-1}$ is conductivity of copper, $\omega = 2\pi f_1$ - circular frequency of acceleration. Substituting numbers for the above parameters of a cylindrical cavity, we find: $\delta$ = 1.8 µ, the internal surface area of the cylindrical cavity with radius $d_{cav} / 2$ = 9 cm and a height $l_{in}$ = 1 cm, isequal to: $S_{in} = \pi d_{cav} (d_{cav} / 2 + l_{in})$ = 565 $cm^2$. The volume of the cavity $V = (\pi d^2_{cav} / 4) * l_{in}$ = 254 $cm^3$. Thus, its quality factor Q is: $Q = V / S_{in} * \delta = 2.5 * 10^3$.

The characteristic impedance of the cylindrical cavity for fashion $E_{010}$ is [7],

$$\rho_s = (\mu/\varepsilon)^{1/2} * l_{in}/d_{cav}, \qquad (7)$$

where $(\mu/\varepsilon)^{1/2}$ is the wave impedance of vacuum, $(\mu/\varepsilon)^{1/2} = 370\ \Omega$, in practical unit system. Substituting the numbers into the formula (7), we find that the characteristic impedance of the cavity $\rho_s$ is equal to $\rho_s = 20\ \Omega$. The shunt resistance of this cavity is equal to: $R_{sr} = \rho_s * Q = 50\ k\Omega$.

If it is necessary to obtain the energy accumulation rate in a set of warm cavities equal to $\Delta W / \Delta l = 20$ MeV / m, then the voltage between the end walls of the cavity at a distance between them equal to $l_{in} = 1$ cm, must be $U = 200$ kV. According to formula (6) it would have been necessary to introduce the power of $P = 4*10^5$ W into the cavity. Then per one meter of the accelerator it would have been required to use the power of 40 MW / m, without taking into account the power transmission into the beam. This power is too large and, therefore, the accelerators consisting of a set of separate short cylindrical warm are not used.

The situation is different for the superconducting accelerators, where the Q-factor of a single cavity is by 7 orders of magnitude higher [6]. The use of superconducting cavities allows one to reduce the power required for the excitation of the cavity by the same orders of the magnitude. Under increasing the Q-factor by 3 orders the power required for the excitation of the cavities will be equal to about 40 kW / m due to superconductivity. The power transferred to the beam is much greater 20 MV / m * 0.7 A = 14 MW / m, so total required high-frequency power will be of about 15 MW / m. This can be considered an acceptable payment for achieving a high (20 MeV / m) rate of proton acceleration.

Parameters of the accelerator consisting of a set of superconducting cavities are shown in Table 3.

Table 3 Parameters of the main part of the accelerator.

| Parameter | Value |
|---|---|
| The frequency of the accelerations f, GHz | 1.3 |
| Average tension of electric field $E_0$, MV/m | 20 |
| High-frequency power, MW | 15 MW/m |
| The Q-factor of cavities | $2.5*10^6$ |
| Tension of the focusing magnetic field, T | 10 |
| Final proton energy, GeV | 1 |

| The length of the accelerator, m | 50 |

**6. Conclusion**

The accelerator can operate together with the multiplying neutron target which is a sub-critical nuclear assembly. This assembly has the neutron multiplication factor of about 20. On the intensity of the neutron flux such an accelerator together with the multiplying target will be better than the European Spallation Source, ESS [8].